\documentclass[a4paper,floatfix,prd,superscriptaddress,reprint,11pt]{article}
\usepackage[utf8]{inputenc}
\usepackage[pdftex]{graphicx}
\usepackage{pos}
\usepackage{fontenc}
\usepackage{times}
\usepackage{graphicx}
\usepackage{appendix}
\usepackage{amsmath}
\usepackage{dsfont}
\usepackage{array}
\usepackage{bm}
\usepackage{multirow}
\usepackage{upgreek}
\usepackage{xcolor}
\usepackage{verbatim}
\usepackage{slashed}
\usepackage{cleveref}
\usepackage{caption}
\usepackage{subcaption}
\usepackage{mathtools}
\usepackage{url}

\usepackage[braket,qm]{qcircuit}
\usepackage{xcolor}

\title{Review on Quantum Computing for Lattice Field Theory}
\ShortTitle{Review on Quantum Computing for Lattice Field Theory}

\author*[a,b]{Lena Funcke}
\author[c]{Tobias Hartung}
\author[d]{Karl Jansen}
\author[d,e]{Stefan Kühn}

\affiliation[a]{Transdisciplinary Research Area ``Building Blocks of Matter and Fundamental Interactions'' (TRA Matter) and Helmholtz Institute for Radiation and Nuclear Physics (HISKP), University of Bonn, Nußallee 14-16, 53115 Bonn, Germany}

\affiliation[b]{Center for Theoretical Physics, Co-Design Center for Quantum Advantage, and NSF AI Institute for Artificial Intelligence and Fundamental Interactions, Massachusetts Institute of Technology, 77 Massachusetts Avenue, Cambridge, MA 02139, USA}

\affiliation[c]{Northeastern University - London, Devon House, St Katharine Docks, London, E1W 1LP, United Kingdom}

\affiliation[d]{Deutsches Elektronen-Synchrotron DESY, Platanenallee 6, 15738 Zeuthen, Germany}

\affiliation[e]{Computation-Based Science and Technology Research Center, The Cyprus Institute, 20 Kavafi Street,
2121 Nicosia, Cyprus}

\emailAdd{lfuncke@uni-bonn.de}
\emailAdd{tobias.hartung@desy.de}
\emailAdd{karl.jansen@desy.de}
\emailAdd{stefan.kuehn@desy.de}

\abstract{In these proceedings, we review recent advances in applying quantum computing to lattice field theory. Quantum computing offers the prospect to simulate lattice field theories in parameter regimes that are largely inaccessible with the conventional Monte Carlo approach, such as the sign-problem afflicted regimes of finite baryon density, topological terms, and out-of-equilibrium dynamics. First proof-of-concept quantum computations of lattice gauge theories in (1+1) dimensions have been accomplished, and first resource-efficient quantum algorithms for lattice gauge theories in (1+1) and (2+1) dimensions have been developed. The path towards quantum computations of (3+1)-dimensional lattice gauge theories, including Lattice QCD, requires many incremental steps of improving both quantum hardware and quantum algorithms. After reviewing these requirements and recent advances, we discuss the main challenges and future directions.\\\\ Preprint number: MIT-CTP/5482}

\FullConference{The 39th International Symposium on Lattice Field Theory, (Lattice 2022)\\
  8-13 August 2022\\
  Hörsaalzentrum Poppelsdorf, Bonn, Germany
}

\begin{document}
\maketitle

\section{Introduction}

Gauge theories are at the heart of our understanding of modern high-energy physics, with the Standard Model of particle physics being arguably the most prominent example. Discretizing gauge theories on a spacetime lattice allows for powerful numerical simulations based on Markov chain Monte Carlo (MCMC) methods. However, despite the great success of MCMC methods, they cease to work in certain parameter regimes due to the infamous sign problem, in particular in the presence of $\theta$-terms, baryon chemical potentials, or out-of-equilibrium dynamics. Various methods have been proposed to alleviate or circumvent this problem, including complex Langevin~(see, e.g., Ref.~\cite{Berger:2019odf}), contour deformations using Lefschitz thimbles~(see, e.g., Ref.~\cite{Cristoforetti:2012su}), as well as approaches based on machine learning~(see, e.g., Ref.~\cite{Boyda:2022nmh}), tensor networks (TN)~(see, e.g., Refs.~\cite{Banuls2018a, Banuls2019}), and quantum computing (see, e.g., Refs.~\cite{Banuls2020,Clemente_Strategies_2022}). Moreover, MCMC methods suffer from critical slowing down, i.e., rapidly growing autocorrelation times when the lattice spacing is decreased, which could be overcome with Hamiltonian-based approaches such as TN or quantum computing.

In these proceedings, we review the current status and future prospects of applying quantum computing to lattice field theory. First quantum computations of gauge 
theories in (1+1) dimensions have already been performed, which demonstrate some of the characteristic features of the Standard Model~(see, e.g., Refs.~\cite{Martinez2016, Yang2020,Mil_2020,A_Rahman_2021,Zhou:2021kdl,Ciavarella:2021lel, Ciavarella:2021nmj, Klco:2019evd, Klco:2018kyo,deJong:2021wsd, Kokail2018, Schweizer:2019, mildenberger_probing_2022, atas2021su2, Atas:2022dqm, Farrell:2022wyt, Farrell:2022vyh,Huffman:2021gsi,Lumia:2021tpu}). Moreover, several resource-efficient formulations of (1+1)- and (2+1)-dimensional gauge theories for quantum computations have already been developed~(see, e.g., Refs.~\cite{Banuls2017,Celi_2020,Kaplan2020,Haase2020,Paulson2020,armon2021photonmediated,Bauer2021}). However, the ambitious goal of quantum computing (3+1)-dimensional phenomena within and beyond the Standard Model is still rather far away, given the currently available Noisy Intermediate-Scale Quantum (NISQ) hardware~\cite{Preskill2018}. To achieve this goal, many incremental steps have to be taken, in particular for improving quantum hardware, quantum algorithms, and quantum error correction.  Furthermore, validating quantum computations is essential for getting reliable results, especially in parameter regimes that are inaccessible with MCMC methods. This validation can be performed using classical Hamiltonian methods, in particular TN-based approaches~\cite{Orus2014a}, for cross-checking the results of quantum computations in the slightly entangled regimes. 

These proceedings are organized as follows. In Sec.~\ref{sec:QCLFT}, we review the computational challenges of MCMC algorithms, comment on classical approaches to address these problems, and discuss why classical computing may (not) be enough. In Sec.~\ref{sec:analog}, we provide an example for outperforming specific classical computations with analog quantum simulators. In Sec.~\ref{sec:QC}, we briefly review the basics of digital quantum computing, introducing the concepts of qubits, quantum circuits, quantum errors, as well as quantum error mitigation and correction. In Sec.~\ref{sec:algorithms}, we review recent advances in developing quantum algorithms for lattice field theory. We discuss hybrid quantum-classical algorithms, quantum circuit design, and methods to implement fermions and gauge fields. We also show a few examples of lattice field theories that have already been simulated on quantum computers. We provide conclusions, discussions, and an outlook in Sec.~\ref{sec:summary}.

\section{Why quantum computing for lattice field theory?}
\label{sec:QCLFT}

\subsection{Computational challenges of MCMC algorithms}
\label{sec:classical}

Classical MCMC simulations usually rely on a Euclidean space-time formulation of lattice field theories and, thus, cannot access real-time dynamics. Therefore, phenomena like the Schwinger effect leading to electron-positron production in strong electric ﬁelds or the out-of-equilibrium dynamics following heavy-ion and proton-collisions at the Relativistic Heavy Ion Collider (RHIC) and the Large Hadron Collider (LHC) cannot be accessed with the method. 
Moreover, the MCMC approach fails at studying strongly-coupled matter at high density. Thus, this approach cannot provide precise numerical results for the QCD equation of state, which is crucial for understanding future experimental data from gravitational-wave observations coming from neutron star collisions, e.g., as observed at the Laser Interferometer Gravitational-Wave Observatory (LIGO).

The origin of these classical computational challenges is the infamous sign problem (see, e.g., Ref.~\cite{Troyer2005}), which is the problem of numerically evaluating the integral of a highly oscillatory function of a large number of variables. Numerical MCMC methods fail here because of the near-cancellation of the positive and negative contributions to the integral, i.e., both have to be integrated to high precision in order for their difference to be obtained with useful accuracy. The number of Monte Carlo sampling points needed to obtain an accurate result rises exponentially with the volume of the system.

Furthermore, MCMC methods are inherently afflicted with autocorrelation times. These autocorelation times grow rapidly when the lattice spacing is decreased, which is a problem called ``critical slowing down''. In some cases, the autocorrelation times can even grow exponentially, see Ref.~\cite{Schaefer:2010hu}, which makes it hard to investigate the continuum limit.

\subsection{Do we really need \textit{quantum} computing?} 

Once we encounter exponentially hard problems of classical algorithms, the first question we may ask is whether quantum algorithms could be used to solve these problems efficiently. This is given for the problem of critical slowing down (see, e.g., Ref.~\cite{Clemente_Strategies_2022}) and also for the sign problem of MCMC methods (see, e.g., Ref~\cite{Banuls2020}). However, the second question we should ask is: Do we really need quantum computing to circumvent these problems, or are there also classical approaches? 

Indeed, there are several classical approaches to reduce or circumvent the obstacles of MCMC methods, in particular the problem of critical slowing down and the sign problem~(see, e.g., Refs.~\cite{Ammon:2015mra,Ammon:2016jap,Boyda:2022nmh,Abbott:2022hkm,Berger:2019odf,Cristoforetti:2012su,Hartung-Jansen-Sarti2022}). For example, approaches for overcoming the problem of critical slowing down have been proposed based on machine learning~(see, e.g., Refs.~\cite{Boyda:2022nmh,Abbott:2022hkm}). Regarding the sign problem, complex Langevin~(see, e.g., Ref.~\cite{Berger:2019odf}), contour
deformations using Lefschitz thimbles (see, e.g., Ref.~\cite{Cristoforetti:2012su}), methods based on machine learning~(see, e.g., Ref.~\cite{Boyda:2022nmh}), and further approaches have been shown to alleviate the sign problem. One of the most promising techniques to completely circumvent the sign problem is the Hamiltonian formulation, which can be efficiently addressed with classical approaches based on TN. This numerical method was originally developed in the context of condensed-matter physics, but has found numerous applications in other fields, including lattice field theory~\cite{Banuls2018a} and even quantum gravity (see, e.g., Refs.~\cite{Dittrich2013,Dittrich2016,Asaduzzaman2020}). 

The TN approach can be used to validate quantum computations in moderately entangled regimes. Therefore, we will review this approach in the following. For details on other methods tackling computational challenges of MCMC simulations, we refer the reader to, e.g., Refs.~\cite{Berger:2019odf, Cristoforetti:2012su, Boyda:2022nmh}.

\subsubsection{Classical methods: example of tensor networks}
\label{sec:TN}

TN states are a specific representation for quantum many-body states, based on the entanglement structure of the state. To illustrate the method, let us consider a generic quantum state $|\psi\rangle$ for $N$ interacting quantum systems. The wave function of the system can be expressed as the sum over all the basis states,
\
\begin{equation}
|\psi\rangle=\sum_{i_1,\ldots i_N=1}^dC_{i_1\ldots i_N}|i_1\rangle\otimes|i_2\rangle\otimes\dots\otimes|i_N\rangle,
\label{eq:psi_C}
\end{equation}
where the states $|i_k\rangle$ form a basis for the $d$-dimensional Hilbert space of the quantum system at site~$k$. The tensor $C_{i_1\ldots i_N}$ that multiplies the many-body basis states has an exponentially large number of complex entries ($d^N$) and, thus, cannot be stored efficiently on a classical computer. The basic idea of TN is to decompose the $C_{i_1\ldots i_N}$ into a network of smaller tensors. A simple example for a family of one-dimensional TN states are Matrix Product States (MPS), which parametrize the coefficients of the wave function as a product of matrices, 
\begin{align}
   |\psi\rangle = \sum_{i_1,i_2,\dots,i_N}^d \text{tr}\left(A^{i_1}_1A^{i_2}_2\cdots A^{i_N}_N\right)|i_1\rangle\otimes|i_2\rangle\otimes\dots\otimes|i_N\rangle.
   \label{eq:psi_MPS}
\end{align}
In the expression above, $A^{i_k}_k$ are complex matrices of size $D\times D$ and $\text{tr}$ denotes the trace. The parameter $D$ is called the bond dimension of the MPS and limits the amount of entanglement that can be present in the ansatz\footnote{Considering the bipartion of the system into two contiguous subsystems, one can show that the von Neumann entropy for the reduced density operator describing a subsystem is upper bounded by $2\log(D)$~\cite{Orus2014a}.}~\cite{Schollwoeck2011,Orus2014a,Bridgeman2017}. In particular, for the example of the MPS ansatz, one has to store $NdD^2$ complex numbers. Thus, provided that $D$ does not grow exponentially with $N$, the representation in terms of a TN is efficient, and it allows for overcoming the exponential scaling in the system size. Results from quantum information theory show that for many physically relevant situations, $D$ does indeed only show a polynomial dependence on $N$~\cite{Vidal2004,Verstraete2006,Hastings2007}. The MPS ansatz can be immediately generalized to higher dimensions~\cite{Verstraete2004b}, and there exist more general TN ansätze for one and higher dimensions~\cite{Shi2006,Vidal2008,Tagliacozzo2009,Evenbly2014}.

Besides being a powerful theoretical tool, the TN formalism allows for an efficient computation of ground states and low-lying excitations. Given a (local) Hamiltonian $H$, one can find a TN approximation for the ground state through variationally minimizing the energy $E=\langle \psi|H|\psi\rangle$ by iteratively updating the tensors~\cite{Schollwoeck2011,Orus2014a,Bridgeman2017}. Subsequently, one can obtain excitations by considering a Hamiltonian projected onto the subspace orthogonal to the ground state. The  ground state is an  eigenstate with vanishing eigenvalue of the projected Hamiltonian, and the first excited state is an eigenstate with energy $E_1$. Given that $E_1 < 0$, which one can always achieve by adding an appropriate constant to $H$, the first excitation corresponds then to the ground state of the projected Hamiltonian, which can be obtained by using the variational algorithm presented above~\cite{Banuls2013}. Moreover, the TN formalism allows to compute the evolution of a quantum system in time, as long as the entanglement in the state during the evolution stays moderate~\cite{Schollwoeck2011,Orus2014a,Bridgeman2017}. In both cases, one can obtain the expectation values of (local) observables $O$ by directly computing $\langle \psi|O|\psi\rangle$. Thus, no MCMC sampling is required, and the TN approach is free from the sign problem. This has been successfully demonstrated for various models, as we will discuss in the next subsection.

\subsubsection{Applying tensor-network methods to lattice field theory}

\begin{figure}[!t]
    \centering
    \subfloat[\centering TN results for (1+1)-dim.\ Schwinger model~\cite{Funcke:2019zna}.]{{\includegraphics[width=.45\linewidth]{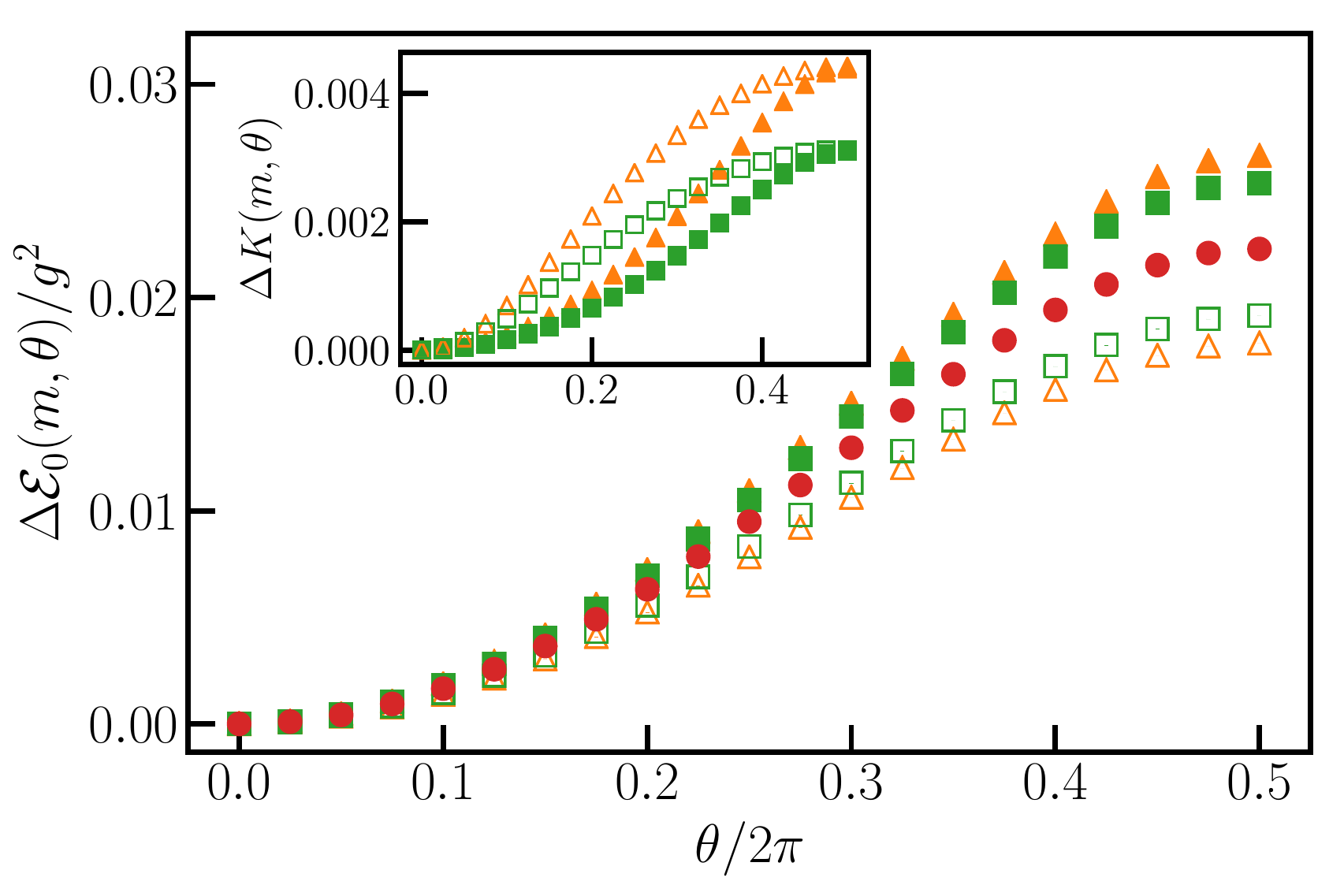} }}
    \qquad
    \subfloat[\centering TN results for (3+1)-dim.\ Lattice QED~\cite{Magnifico:2020bqt}.]{{\includegraphics[width=.40\linewidth]{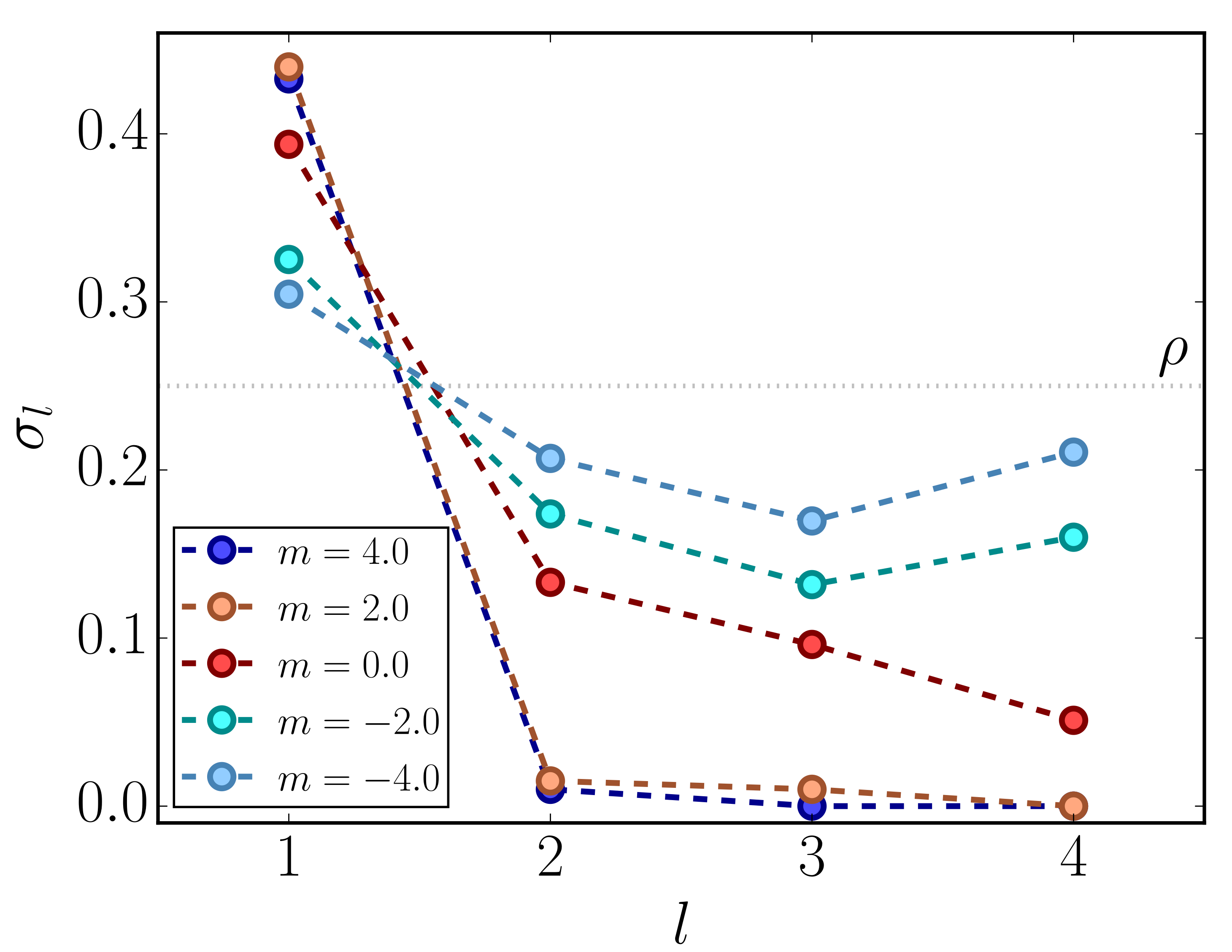} }}
    \caption{(a) Ground-state energy density of the Schwinger model as a function of the $\theta$-parameter, for the bare mass $m/g=0.07$ (filled markers) and for $m/g=-0.07$ and $\theta\to\theta +\pi$ (open markers). The orange triangles (green squares) correspond to finite-lattice data with $x=80$ ($x=160$), where $x\equiv 1/(ag)^2$, $a$ is the lattice spacing, and $g$ is the coupling. The red dots represent the results after extrapolating the finite-lattice data to the continuum.  Inset: Absolute value of the deviations between the lattice and continuum data,
    which demonstrates the lattice distortion of the anomaly equation. Figure and caption adapted from Ref.~\cite{Funcke:2019zna}. (b) Surface charge density $\sigma_l$ of (3+1)-dimensional Lattice QED on a cube whose faces are at distance $l$ from the boundaries of the lattice with linear size $L = 8$, for different values of the fermion mass $m$. The system is in the global symmetry sector with $Q = 128$ positive charges (ﬁnite density $\rho = 1/4$). Figure reprinted with permission from Ref.~\cite{Magnifico:2020bqt}, \url{https://doi.org/10.1038/s41467-021-23646-3}, under the \href{https://creativecommons.org/licenses/by/4.0/}{Creative Commons Attribution 4.0 International License}. }
    \label{fig:TN}
\end{figure}

Investigations of lattice field theories with TN methods have progressed substantially in recent years. 
For example, the MPS approach has been used to explore quantum link models, $\mathds{Z}_n$-QED models, and non-Abelian gauge models. Moreover, the spectrum of the Schwinger model has been computed using MPS, and the model has been studied at non-zero chemical potential, non-zero $\theta$-term, non-zero temperature, and for real-time problems.  Going beyond MPS, also TN renormalization techniques~\cite{Levin2007,Yang2015,Evenbly2015b} have been used to investigate various gauge theories in (1+1) dimensions, as well as the CP(1) model with a $\theta$-term. See Refs.~\cite{Banuls2018a,Banuls2019} for reviews and Refs.~\cite{Kuramashi2018,Silvi2019,Magnifico:2019ulp,Funcke:2022uwc,Funcke:2019zna,Funcke:2021glr,Angelides:2022pah,Nakayama:2021iyp,Kuramashi:2018mmi,Felser:2019xyv,Magnifico:2020bqt} for more recent studies not included in Refs.~\cite{Banuls2018a, Banuls2019}. In particular, recently there have been first TN studies of (2+1)-dimensional gauge theories~\cite{Kuramashi:2018mmi,Felser:2019xyv} and the first TN study of Lattice QED in (3+1)-dimensions at finite density~\cite{Magnifico:2020bqt}.

In Fig.~\ref{fig:TN}, we show two examples of applying TN methods to overcoming the sign problem, which arises due to a $\theta$-term [Fig.~\ref{fig:TN}(a)] or due to finite density [Fig.~\ref{fig:TN}(b)]. Figure~\ref{fig:TN}(a) shows results obtained in Ref.~\cite{Funcke:2019zna}, where the topological vacuum structure of the Schwinger model with a $\theta$-term was studied. The Schwinger model shares many similarities with (3+1)-dimensional QCD and thus serves as benchmark model for testing new numerical techniques aimed at Lattice QCD applications.  
Using MPS, Ref.~\cite{Funcke:2019zna} simulated parameter regimes of the Schwinger model for which both perturbation theory and MCMC methods break down. In particular, Ref.~\cite{Funcke:2019zna} quantified the lattice distortion of the quantum anomaly equation that maps negative to positive fermion masses, $m\to -m$, by shifting the $\theta$-parameter, $\theta\to \theta +\pi$, see Fig.~\ref{fig:TN}(a). 

Figure~\ref{fig:TN}(b) shows results obtained in Ref.~\cite{Magnifico:2020bqt}, where (3+1)-dimensional Lattice QED was simulated at finite charge density.
Using tree TN, Ref.~\cite{Magnifico:2020bqt} computed the surface charge density $\sigma_l=\frac{1}{A(l)}\sum_{{\color{black} \mathbf{x}}\in A(l)}\langle \hat \psi_{{\color{black} \mathbf{x}}}^{\dagger} \hat \psi_{{\color{black} \mathbf{x}}}\rangle$, where ${\color{black}\mathbf{x}}  \equiv \left ( i,j,k \right )$ for $0 \leq i,j,k \leq  L-1$ labels the sites of the lattice, $L$ is the lattice size, $\hat\psi_{{\color{black}\mathbf{x}}}$ is the staggered spinless fermion field, and $A(l)$ contains only sites sitting at a particular lattice distance $l$ from the closest boundary. These first (3+1)-dimensional TN calculations are a milestone for classically tackling the sign problem, but many algorithmic advances are still required to apply TN methods to (3+1)-dimensional Lattice QCD.

\subsection{Why is(n't) classical computing enough?}

Although alternative classical methods are able to overcome or alleviate some of the limitations of the conventional MCMC approach, so far it has not been possible to fully access the regimes that cannot be addressed with MCMC simulations. Quantum computers offer the prospect to overcome these restrictions. In the following, we discuss these limitations focusing on the TN approach.

As outlined above, the success of the TN approach crucially relies on the fact that most physically relevant states only carry little entanglement and, thus, can be described with moderate tensor size. In particular, this encompasses the ground states of Hamiltonians with local interactions of finite strength and a nonvanishing energy gap in the thermodynamic limit, which are believed to be efficiently described by TN states\footnote{In fact, this can be proven for systems in (1+1) dimensions, see Ref.~\cite{Hastings2007}. For higher dimensions, it is conjectured that ground states of local gapped Hamiltonians are efficiently described by TN states.}. However, there exist certain situations where the amount of entanglement can be prohibitively large, preventing an efficient TN description. A prominent example are out-of-equilibrium dynamics following a global quench, during which the entanglement in the system can grow linearly in time~\cite{Calabrese2005,Schuch2008a}. This would require the tensor size to grow exponentially to maintain a faithful representation of the wave function of the system. In these situations, TN simulations only allow for accessing short time scales before the exponential growth of the tensor size renders the computation infeasible.

Moreover, even in situations where an efficient TN description is possible, the computational cost of TN algorithms, despite being polynomial in the tensor size, might prevent calculations in practice. For the MPS introduced above, the leading-order computational cost of the variational ground-state search and the simulation of time evolution scales as $\mathcal{O}(D^3)$~\cite{Schollwoeck2011,Orus2014a,Bridgeman2017}. For its two-dimensional generalization, the projected entangled pair states (PEPS)~\cite{Verstraete2004b}, the algorithms for variational ground-state optimization can scale up to $\mathcal{O}(D^{10})$~\cite{Lubasch2014}. While this is still polynomial in the bond dimension, this significantly limits the values of the bond dimensions that can be reached in practical calculations. Although there have been recent developments in designing TN ansätze suited to tackle (2+1)- and (3+1)-dimensional lattice gauge models with first successful computations~\cite{Kuramashi:2018mmi,Felser:2019xyv,Magnifico:2020bqt}, extending the success of TN computations to higher dimensions is not an immediate task. Quantum computing and quantum simulation might offer an alternative route to overcome these limitations, as we discuss in the following section.

\section{Outperforming tensor-network methods with analog quantum simulators}
\label{sec:analog}

\begin{figure}[!htb]
    \centering
    \includegraphics[width=.6\linewidth]{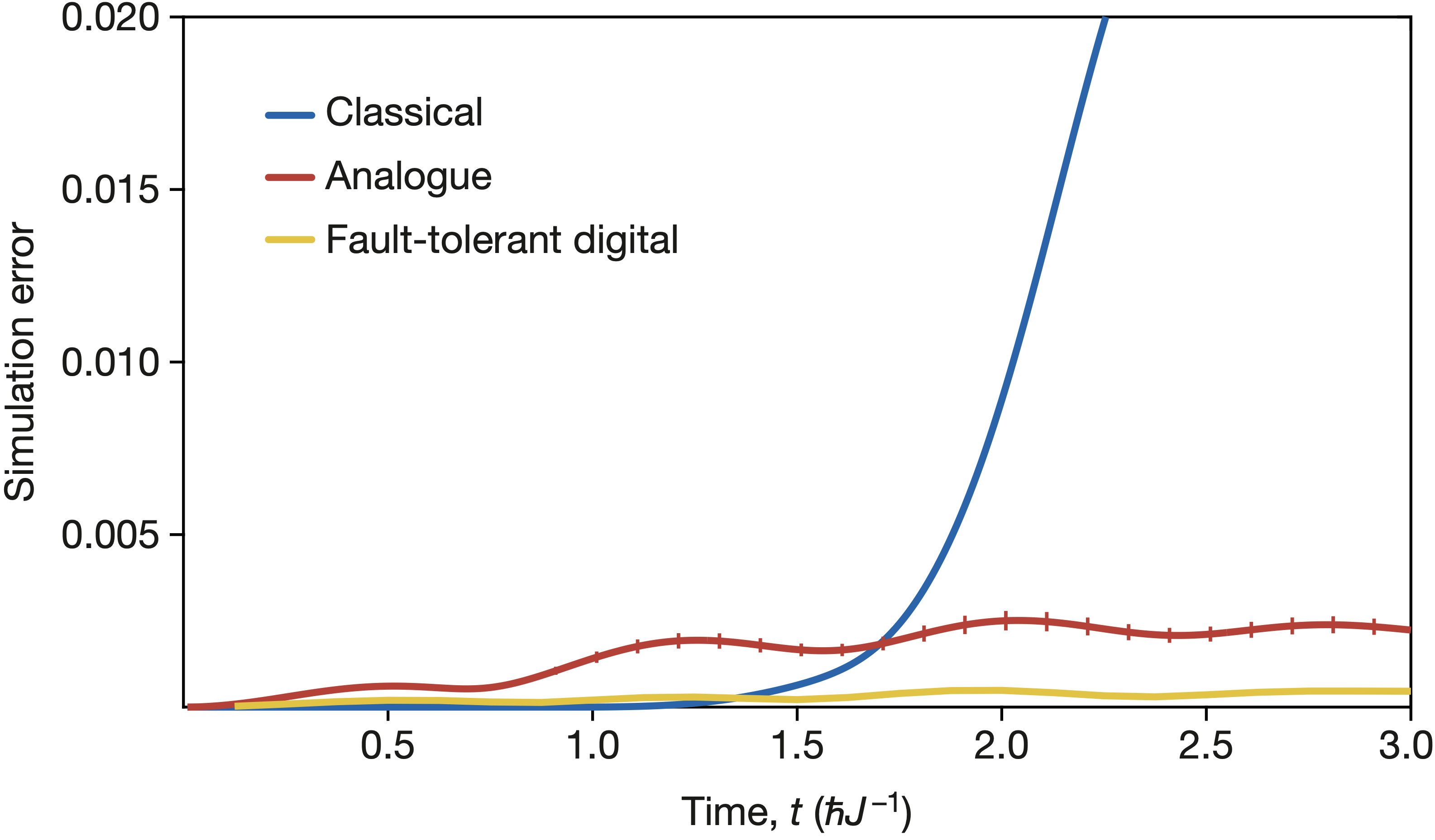}
    \caption{Quantum advantage of quantum simulation over classical simulation, illustrated by errors in single-particle correlations arising for quench dynamics of a (1+1)-dimensional Bose-Hubbard model, see Eq.~\eqref{eq:BoseHubbard}, for $U/J=1$ and 20 sites. 
    The blue line is a single-site time-dependent variational principle time evolution using MPS with bond dimension $D=64$, as a demonstration of typical truncation errors from these classical methods relative to exact calculations, which are necessary for longer times and system sizes. The red line is an analog simulation with calibration errors of 1\%. The yellow line is the digital simulation with a second-order Trotter decomposition with a time step $J\delta t=\hbar/8$. Figure reprinted from Ref.~\cite{Daley:2022eja}, \url{https://doi.org/10.1038/s41586-022-04940-6}, with permission from Springer Nature.}
    \label{fig:TN-vs-QC}
\end{figure}

Already in 2012, Ref.~\cite{Trotzky_2012} demonstrated that analog quantum simulations can outperform classical simulations for the case of real-time evolution, in particular, for simulating the relaxation towards equilibrium in an isolated strongly correlated (1+1)-dimensional Bose gas: ``the controlled [quantum] dynamics runs for longer times than present classical algorithms can keep track of''~\cite{Trotzky_2012}. Even though this quote essentially translates to the nowadays popular term of ``quantum advantage'', this term has not been used back in 2012; only ten years later, Ref.~\cite{Daley:2022eja} argued that a ``practical quantum advantage'' has already been achieved using analog quantum simulators. The reason for the failure of the classical simulation methods, such as the MPS-based methods used in Ref.~\cite{Trotzky_2012}, is that these classical methods suffer from an extensive increase in entanglement entropy, as outlined in the previous section. This limits the relaxation times accessible in the classical calculations, as shown in Fig.~\ref{fig:TN-vs-QC}.

\begin{figure}[!t]
    \centering
    \includegraphics[width=.8\linewidth]{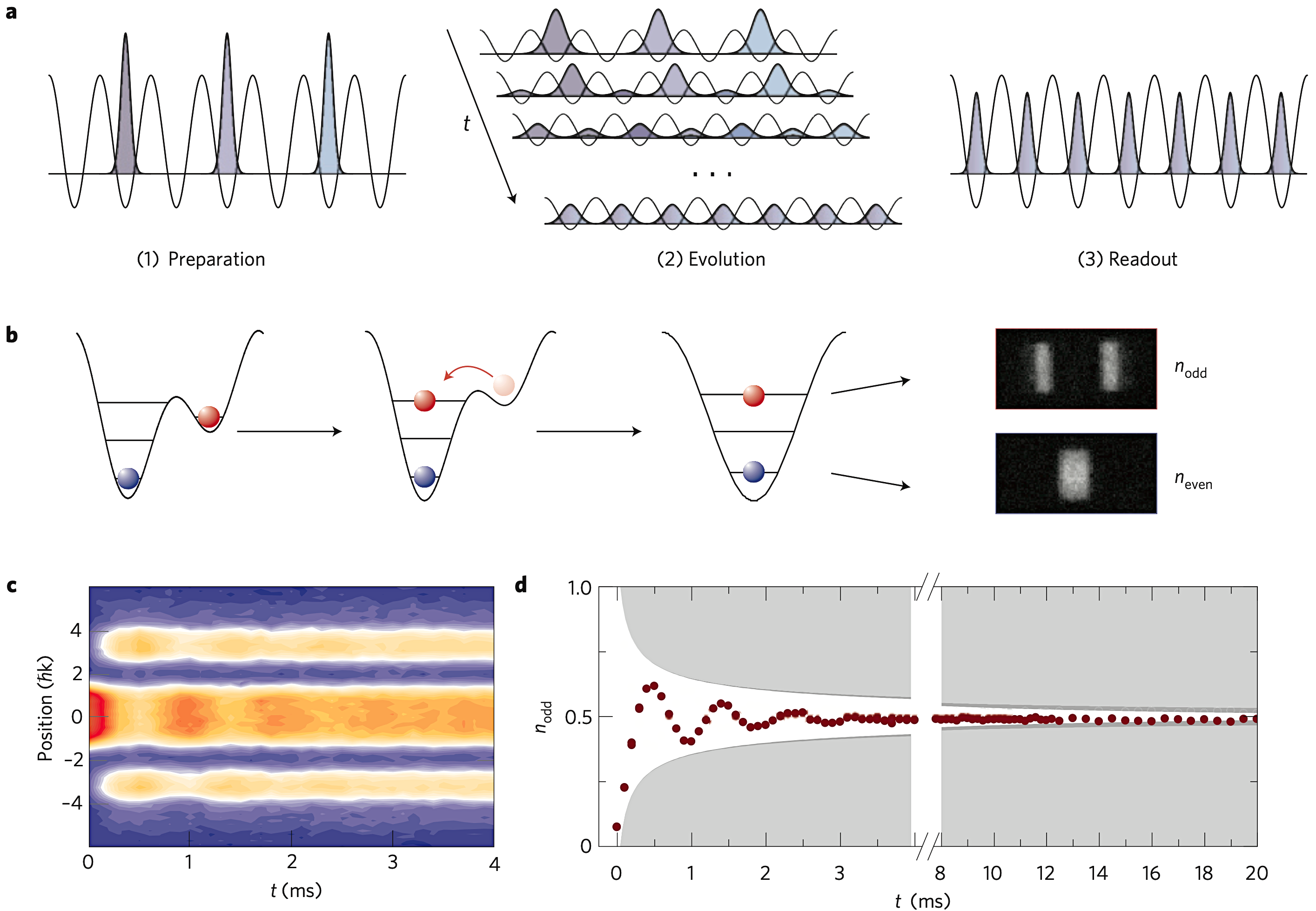}
    \caption{Analog quantum simulation of the (1+1)-dimensional Bose-Hubbard model, see Eq.~\eqref{eq:BoseHubbard}. (a)~Concept of the experiment: after having prepared the density wave $|\psi(t=0)\rangle$ ({\it i}), the lattice depth was rapidly reduced to enable tunneling ({\it ii}). Finally, the properties of the evolved state were read out after all tunneling was suppressed again ({\it iii}). (b) Even-odd resolved detection: particles on sites with odd index were brought to a higher Bloch band. A subsequent band-mapping sequence was used to reveal the odd- and even-site populations. (c) Integrated band-mapping profiles versus relaxation time $t$ for $h/(4J) \simeq 0.9\,{\rm ms}$, $U/J = 5.16(7)$ and $K/J \simeq 9 \times 10^{-3}$. (d) Odd-site density extracted from the raw data shown in c. The shaded area marks the envelope for free bosons (light grey) and including inhomogeneities of the Hubbard parameters in the experimental system (dark grey). Figure reprinted from Ref.~\cite{Trotzky_2012}, \url{https://doi.org/10.1038/s41586-022-04940-6}, with permission from Springer Nature.}
    \label{fig:BoseHubbard}
\end{figure}

In Ref.~\cite{Trotzky_2012}, the quantum theory is a (1+1)-dimensional chain of lattice sites coupled by a tunnel coupling $J$ and filled with repulsively interacting bosonic particles. In the tight-binding approximation, the Hamiltonian takes the form of a (1+1)-dimensional  Bose-Hubbard model,
\begin{eqnarray}
  H_{\rm BH_{(1+1)D}}  &=& \sum_j \biggl[-J\left(\hat a^\dagger_j \hat a^{\phantom{\dagger}}_{j+1} +\mbox{h.c.}\right)
     +\frac{U}{2}   \hat n_j(\hat n_j -1) + \frac{K}{2}  \hat n_j j^2\biggr]\,,
     \label{eq:BoseHubbard}
\end{eqnarray}
where $\hat a_{j}$ annihilates a particle on site $j$, $\hat n_{j} = \hat a_{j}^\dagger \hat a^{\phantom{\dagger}}_{j}$ corresponds to the number of atoms on site $j$, and $U$ is the on-site interaction energy. The parameter $K = m \omega^2 d^2$ ($m$ is the particle mass, $d$ the lattice spacing) describes an external harmonic trap with trapping frequency $\omega \simeq 2\pi \times 61\,{\rm Hz}$. Details of the analog quantum simulation of the model in Eq.~\eqref{eq:BoseHubbard} are shown in Fig.~\ref{fig:BoseHubbard}.

Here, we note that the ``practical quantum advantage'' has only been demonstrated with analog quantum simulators~(see Ref.~\cite{Daley:2022eja} and references therein), not with digital quantum computers. In the following, we would like to briefly comment on the differences between these two different types of quantum technology. With an analog quantum simulator, one uses a controllable quantum system to emulate the behavior of another quantum system while exploiting quantum effects. Thus, analog quantum simulators perform continuous time evolution and usually are non-universal. In contrast, digital quantum computers are universal and realize unitary operations via a set of universal logical quantum gates acting onto qubits. Thus, the time evolution has to be decomposed in discrete steps. The universality of digital quantum computers implies that they will eventually be more relevant for the ambitious goal to quantum compute higher-dimensional lattice gauge theories, in particular (3+1)-dimensional Lattice QCD. Thus, in the remaining part of these proceedings, we will only focus on the latter type of quantum technology, namely the digital quantum computers.

\section{Basics of digital quantum computing}
\label{sec:QC}

Quantum computers offer the prospect to outperform classical computers in a variety of tasks ranging from cryptography to machine learning to combinatorial optimization problems. In particular, the potential to efficiently simulate quantum many-body systems makes quantum computers a promising tool for solving quantum many-body problems in physics, chemistry, and beyond, such as the sign problem. In the following, we will review the essential features of digital gate-based quantum computing, which are quantum bits, quantum circuits, and quantum errors. 

\subsection{Quantum bits} 

In classical computing, information is stored and processed as bits which can take definite binary values, 0 or 1. In analogy, quantum bits (qubits) are quantum systems described by a two-dimensional Hilbert space with the basis states $\{|0 \rangle, |1 \rangle\}$. The laws of quantum mechanics allow for creating superpositions of the two basis states, and a general single-qubit state $|\psi\rangle$ can be described by the linear combination
\begin{equation}
|\psi \rangle = \alpha |0 \rangle + \beta |1 \rangle,\quad  |0 \rangle = \begin{pmatrix} 1\\0 \end{pmatrix}, \quad | 1 \rangle = \begin{pmatrix} 0\\1 \end{pmatrix}, 
\end{equation}
where $\alpha$ and $\beta$ are complex numbers constrained by $| \alpha |^2 + | \beta |^2 = 1$ for the state to be normalized. When applying a projective measurement to the qubit in the computational basis, the probability of outcome $|0 \rangle $ is $| \alpha |^2$ and the probability of outcome $|1 \rangle $ is $| \beta |^2$. In particular, the possibility to create superpositions allows for having $N$ qubits in a superposition state of all $2^N$ possible basis states and, thus, to encode an exponentially large Hilbert space efficiently.

How many qubits do current quantum computers have? We have recently entered the Noisy Intermediate-Scale Quantum (NISQ)~\cite{Preskill2018} era, defined as the era of $\mathcal{O}(10-100)$ noisy qubits of digital quantum computers. For the example of IBM-Q quantum devices, the number of qubits has roughly doubled every year, from 27 in 2019 to 433 in 2022, and this journey is planned to be continued with 1121 qubits in 2023 and 4158 qubits in 2025~(see, e.g., Ref.~\cite{Ball_2021}). For the example of Google quantum devices, the aim is to build a ``useful, error-corrected quantum computer''~\cite{Google-Roadmpap2021} by the end of the decade, which should contain 1,000 fault-tolerant logical qubits (see Sec.~\ref{sec:correction} for details on error correction), corresponding to 1,000,000 physical qubits.

\subsection{Quantum computations}
\label{sec:circuit}

The prevailing model of quantum computation describes the computation in terms of a quantum circuit, which consist of three stages: (i) Initialization of all $N$ qubits in the $|0\rangle$ state (denoted by $|0\rangle^{\otimes N}$). (ii) Quantum gates, which represent unitary transformations. (iii) Measurements in the computational basis, on some or all of the qubits.

Quantum computations can be conveniently visualized with a quantum circuit diagram, in which time flows from left to right. Each single line represents a quantum wire, and each double-line represents a classical bit. The items on the quantum wires indicate operations performed on the qubits, such as gate operations or measurements. For example, let us consider a three-qubit circuit
\begin{equation}
\begin{split}
\Qcircuit @C=1em @R=.7em {
    \lstick{\ket{0}} & \gate{R_x(\theta_0)} & \gate{H} & \ctrl{1} & \meter & \cw & \cw \\
    \lstick{\ket{0}} & \gate{R_y(\theta_1)} & \ctrl{1} &  \targ  &  \meter & \cw & \cw \\
    \lstick{\ket{0}} & \qw & \targ & \qw & \meter & \cw & \cw
    }         
\end{split}
\label{eq:circ}
\end{equation}
with parametric rotation gates $R_x(\theta_0)=\exp(-i\frac{\theta_0}{2}\sigma_x)$ and $R_y(\theta_1)=\exp(-i\frac{\theta_1}{2}\sigma_y)$, a Hadamard gate $H$, and two CNOT gates for creating entanglement. The second stage of this circuit is given by the unitary operation
\begin{equation}
(\text{CNOT}\otimes \mathds{1}_2)\times (H\otimes \text{CNOT})\times (R_x(\theta_0)\otimes R_y(\theta_1) \otimes \mathds{1}_2),
\end{equation}
because the composition across wires is achieved by a tensor product, and the composition along wires is achieved by a matrix product. The third stage is a projective measurement in the $\sigma_z$-basis $\{|0\rangle,|1\rangle\}$ yielding classical bits. All these three stages of the quantum computation are affected by quantum noise, as we discuss in the following section.

\subsection{Quantum errors}
\label{sec:errors}

\subsubsection{Which types of errors occur on quantum computers?}

Current intermediate-scale quantum devices suffer from a considerable level of noise, including gate errors, measurement errors, and thermal relaxation errors. For example, there can be bit-flip measurement errors in the last stage of the quantum computation (see Sec.~\ref{sec:circuit}). Due to these bit flips, one erroneously reads out a measurement outcome as $0$ given it was actually $1$, or vice versa. Another example is thermal relaxation errors affecting the qubits. In general, errors can be represented by quantum channels and turn an originally pure state, described by the density operator $\op{\psi}{\psi}$, into a genuinely mixed state described by a density operator $\rho$. The evolution of a quantum state $\rho$ affected by thermal relaxation errors for some time $t$ can, e.g., be expressed as
\begin{equation}
\rho\to\rho'=\sum_{k=1}^2E_k\rho E_k^\dagger, \quad E_1=\begin{bmatrix} 1 & 0 \\ 0 & \sqrt{1-\lambda}\end{bmatrix}, \quad E_2=\begin{bmatrix} 0 & \sqrt{\lambda} \\ 0 & 0 \end{bmatrix},
\end{equation}
where $\lambda=1-\exp(-t/T_1)$. Thus, given a state prepared in $|1\rangle$, the probability that the state is correctly measured in $|1\rangle$ and has not decayed into the state $|0\rangle$ is given by $p(t)= \exp(-t/T_1)$, where $T_1\sim 100\mu$s for current superconducting quantum devices. Although this limits the depth of the circuits that can be executed faithfully, Noisy Intermediate-Scale Quantum (NISQ) devices are already able to exceed the capabilities of classical computes in specific cases. Error mitigation and especially error correction remain one of the major challenges of quantum computing, whose successful implementation is crucial for achieving quantum advantage for large-scale problems relevant in physics and chemistry.

\subsubsection{Error correction versus error mitigation}
\label{sec:correction}

Quantum computers can in principle be made fault-tolerant, according to the threshold theorem for quantum computing (called the ``quantum threshold theorem'')~\cite{Shor_1996,Kitaev:1997wr,Knill:1997mr,Aharonov:1999ei}, which is analogous to the threshold theorem for classical computing by von Neumann~\cite{Neumann_1956}. Fault-tolerance means that quantum error correction techniques can suppress the logical error rate on quantum computers to \textit{arbitrarily} small levels, given that the quantum computer has a physical error rate below a certain threshold. There are many error correction schemes on the market, such as the bit-flip code~\cite{Peres_1985}, the Shor code~\cite{Shor_1995}, the surface code~\cite{Kitaev:1997wr}, and the GKP code~\cite{Gottesman:2000di}. In general, realizing fault-tolerant quantum computation requires two ingredients: (i)~quantum errors below a certain threshold and (ii)~additional (``physical'') qubits to realize the quantum error correction scheme and successfully encode the information of a fault-tolerant (``logical'') qubit. For example, if a quantum computer has a depolarizing error probability below the threshold of 0.1\%, the surface code would require more than 1000 physical qubits per logical qubit~\cite{Campbell_2017}.

Given the current qubit numbers of $\mathcal{O}(10-100)$ and quantum error rates of up to $\mathcal{O}(1\%)$ for gate errors and up to $\mathcal{O}(10\%)$ for measurement errors~\cite{Tannu:2019}, fault-tolerant quantum computation is still far away. Thus, as a near-term solution for NISQ devices, one needs to employ quantum error mitigation instead of quantum error correction. The general concept behind quantum error mitigation is to use a low-overhead method to alleviate the effects of quantum noise and thereby obtain more reliable estimates, e.g., for expectation values of observables. For instance, this can be done by altering the quantum circuit executed on the quantum device, by post-processing the data collected from the device, by measuring modified operators, or by using combinations thereof. A few examples of error mitigation techniques are zero-noise extrapolation, randomized compiling, quasi-probability decomposition, and operator rescaling methods (see, e.g., Refs.~\cite{Kandala:2017,Li:2017,Temme:2017,Endo:2018,Endo:2019,Kandala:2019,Tannu:2019,YeterAydeniz2019, YeterAydeniz2020, chen2021exponential,Cramer_2016,Funcke2020}). 

\subsubsection{Example of error mitigation: zero-noise extrapolation for lattice theory application}

\begin{figure}[h]
    \centering
\includegraphics[width=.6\linewidth]{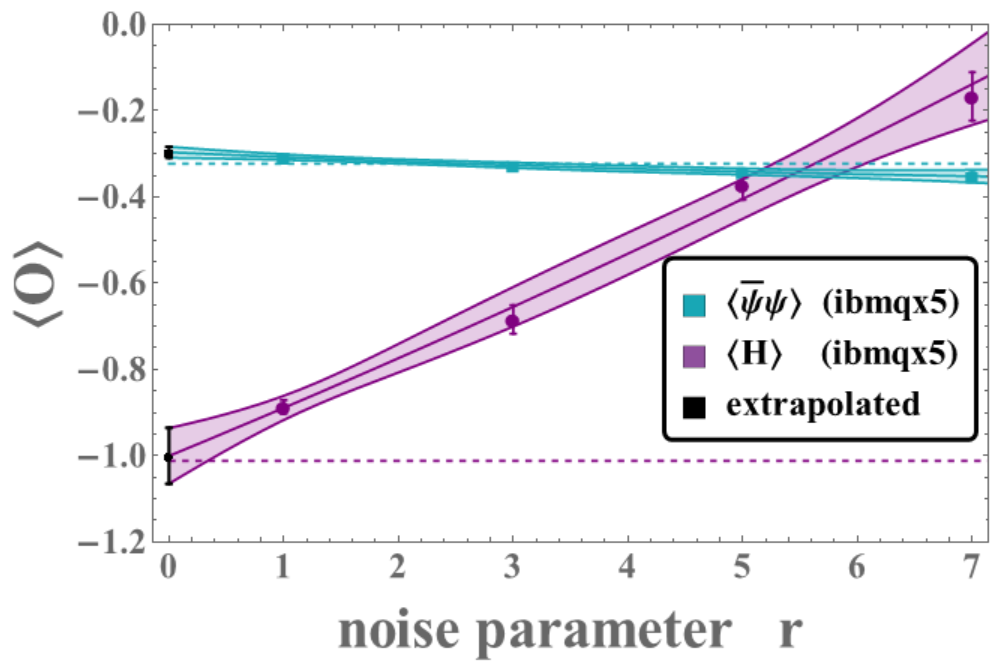}
\caption{Zero-noise extrapolation for hybrid quantum-classical computation of Schwinger model dynamics. Ground-state energy $\langle H\rangle$ and chiral condensate $\langle\bar{\psi}\psi\rangle$ (purple and blue, respectively) as a function of the noise parameter $r$.
The points at $r = 0$ (black) have been quadratically extrapolated. The horizontal dashed lines indicate the exact values. Figure reprinted with permission from Ref.~\cite{Klco:2018kyo}, \url{https://doi.org/10.1103/PhysRevA.98.032331}, copyright (2018) by the Americal Physical Society. }
\label{fig:ZNE}
\end{figure}

Let us briefly discuss one example of error mitigation, i.e., zero-noise extrapolation~\cite{Temme:2017}, including its application to a lattice field theory simulation on a noisy quantum computer~\cite{Klco:2018kyo}. The basic idea behind zero-noise extrapolation is that one cannot easily \textit{reduce} the noise of a quantum circuit, but one can easily \textit{add} noise to a quantum circuit. For example, one might add a pair of CNOT gates behind every CNOT gate in a quantum circuit. In the perfect, noise-free case, this should not alter the quantum computation due to $({\rm CNOT})^2=\mathds{1}$. However, for noisy CNOT gates we have $({\rm CNOT})^2\neq\mathds{1}$, and adding an increasingly large number of CNOT pairs will result in an increasingly large error of the quantum computation. Thus, we may introduce a noise parameter $r$, where $r-1$ is the number of additional CNOT gates inserted at each location of a CNOT gate in the original quantum circuit. Performing the quantum computation at different values of $r$ and subsequently extrapolating to $r\to 0$ therefore offers a way to reduce the systematic bias due to CNOT gate noise. Figure~\ref{fig:ZNE} shows an example of how this method can be applied to lattice field theory, in particular a hybrid quantum-classical computation of Schwinger model dynamics~\cite{Klco:2018kyo} (for another example, see Ref.~\cite{Huffman:2021gsi}). The algorithm behind this hybrid quantum-classical computation will be explained in the following, see Sec.~\ref{sec:VQE}.

\section{Quantum algorithms for lattice field theory}
\label{sec:algorithms}

\subsection{Hybrid quantum-classical algorithms}

\subsubsection{Variational quantum eigensolver for computing ground states}
\label{sec:VQE}

In the context of quantum many-body systems, a promising approach for exploiting NISQ devices is the use of hybrid quantum-classical algorithms, such as the variational quantum eigensolver (VQE)~\cite{Peruzzo2014,McClean2016}. This algorithm makes use of a feedback loop between a classical computer and a quantum coprocessor, where the latter is used to efficiently evaluate the cost function
\begin{equation}
C(\vec{\theta})=\langle\psi(\vec{\theta})|H|\psi(\vec{\theta})\rangle
\end{equation} 
for a given set of variational parameters $\vec{\theta}$. Here, the quantum state $|\psi(\vec{\theta})\rangle$ is realized by a parametric quantum circuit, such as the exemplary circuit in Eq.~\eqref{eq:circ}, where the variational parameters are the rotational angles $\theta_1$ and $\theta_2$. Provided that the ansatz for $|\psi(\vec{\theta})\rangle$ is expressive enough, the minimum for $C(\vec{\theta})$ is obtained for the ground state of the problem Hamiltonian $H$. In the quantum-classical feedback loop, the parameters $\vec{\theta}$ are optimized on a classical computer based on the measurement outcome obtained from the quantum coprocessor.

\subsubsection{Variational quantum deflation for computing excited states}
\label{sec:VQD}

Using an extension of the VQE algorithm described in Sec.~\ref{sec:VQE}, which is called the variational quantum deflation (VQD) algorithm~\cite{Jones_2019,Higgott2019variationalquantum},  one can compute the mass gaps of a lattice field theory. The goal is to estimate the energy of the $k$-th excited state $E_k$ by penalizing the solutions of the lowest excited states. This can be done through minimizing the cost function
\begin{equation}
C(\vec{\theta}_k)=\langle\psi(\vec{\theta}_k)|H|\psi(\vec{\theta}_k)\rangle + \sum_{i=0}^{k-1}\beta_i {|\langle \psi(\vec{\theta}_k)|\psi(\vec{\theta}_i^*)\rangle|}^2,
\label{eq:VQD}
\end{equation}
where $\vec{\theta}_i^*$ are the optimal parameters for the $i$-th excited state and $\beta_i$ are real-valued coefficients, which must be larger than the mass gaps $E_k-E_i$. Thus, one minimizes the variational energy $E(\vec{\theta}_k)$ [the first term in Eq.~\eqref{eq:VQD}] with the constraint [the second term in Eq.~\eqref{eq:VQD}] that the state $|\psi(\vec{\theta}_k)\rangle$ must be orthogonal to the previous $k$ states. 
In order to compute the mass gap, for example, between the ground state $E_0$ and first excited state $E_1$, one follows three steps:
\begin{itemize}
\item Perform the VQE and obtain optimal parameters and an approximate ground state 
$|\psi(\theta_0^*)\rangle$,
\item For the energy $E_1$ of the first excited state, define the Hamiltonian: $H_1 = H + \beta_1 |\psi(\theta_0^*)\rangle\langle\psi(\theta_0^*)|$.
\item Perform the VQE with 
the Hamiltonian $H_1$ to find an approximate first excited state $|\psi(\theta_1^*)\rangle$.
\end{itemize}

It has been experimentally demonstrated~(see, e.g., Refs.~\cite{Banuls2020,klco2021standard, zohar2021quantum} for reviews) that VQE and VQD allow for finding both the ground state and low-lying excitations of low-dimensional benchmark models relevant for particle physics, condensed matter physics, as well as quantum chemistry. To reach applicability in higher dimensions and for larger system sizes, the development of resource-efficient quantum algorithms is crucial. This research area, including the implementation of (gauge) symmetries, has overlap with classical algorithm development, e.g., in TN and machine learning. In particular, the VQE and VQD algorithms are similar to variational TN algorithms (see Sec.~\ref{sec:TN}), now realizing the quantum state $|\psi(\vec{\theta})\rangle$ by a quantum circuit [see Eq.~\eqref{eq:circ}] instead of a tensor network [see Eq.~\eqref{eq:psi_MPS}], thus replacing the tensor parameters with gate parameters.

\subsection{How to design optimal quantum circuits for hybrid quantum-classical algorithms?}
\label{sec:DEA}

\begin{figure}[!b]
    \centering
    \includegraphics[width=\textwidth]{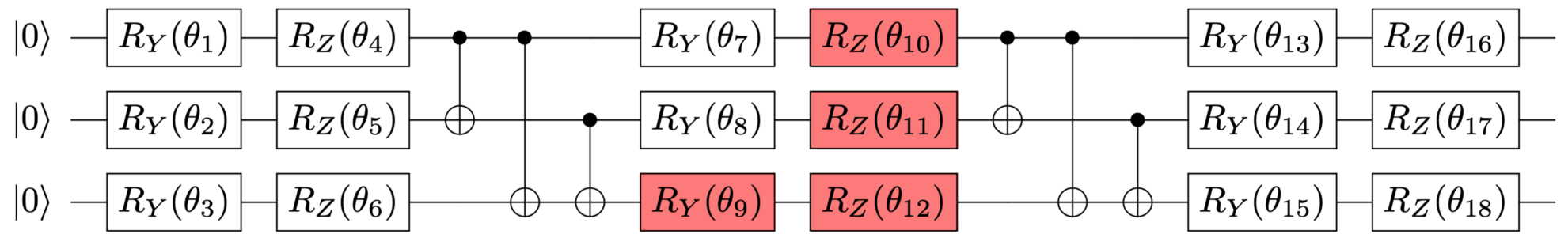}
    \caption{QISKIT's \texttt{EfficientSU2} 2-local circuit with 2 layers. The coloured 
    gates are the gates that can be removed after applying the dimensional expressivity analysis. Figure and caption adapted from Ref.~\cite{Funcke_2021}.}
\label{fig_qiskitsu2}
\end{figure}

Parametric quantum circuits (see Fig.~\ref{fig_qiskitsu2} for an example) are at the heart of variational quantum algorithms (see Sec.~\ref{sec:VQE}). For efficient quantum computation, it is essential to design optimal quantum circuits for obtaining the low-lying energy spectrum of a given problem Hamiltonian. On the one hand, the quantum circuit should contain sufficiently many parametric quantum gates to express the desired solution. On the other hand, the number of quantum gates should be minimal, in order to reduce the noise (see Sec.~\ref{sec:errors}) in the quantum computation. 

There have been many attempts to efficiently design quantum circuits, in particular for lattice field theory applications with physical symmetries (see, e.g., Refs.~\cite{Klco:2018kyo, Klco:2019evd,Martinez2016,Ciavarella:2021lel,Ciavarella:2021nmj,Schweizer:2019,Yang2020,Mil2020,deJong:2021wsd,Nguyen:2021hyk,atas2021su2,Funcke_2021,funcke2021bestapproximation}). One example for a generic method to design minimal and maximally expressive quantum circuits was provided in Refs.~\cite{Funcke_2021,funcke2021bestapproximation}, which was called a dimensional expressivity analysis. This analysis offers a practical, systematic way to optimize a given quantum circuit by removing redundant parameters, incorporating physical symmetries, removing unwanted symmetries, and testing whether the quantum circuit is sufficiently expressive.

\begin{figure}[!b]
    \centering
    \includegraphics[width=2.5in]{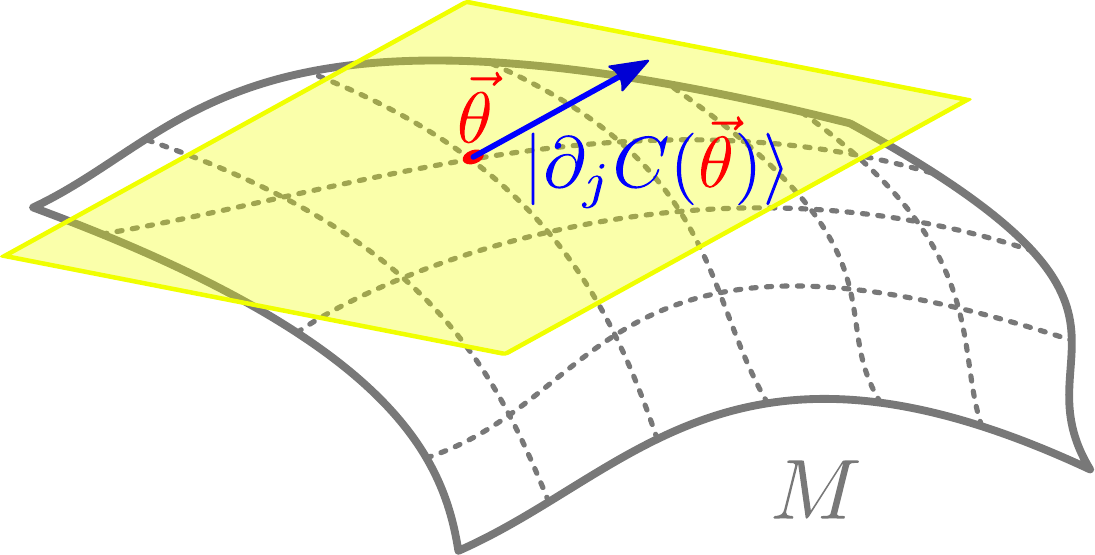}
    \caption{Illustration of the tangent space (illustrated in yellow) of the manifold $M$ (illustrated in grey) of states $|C(\vec{\theta})\rangle$ that a given parametric quantum circuit can reach, in the point $\vec{\theta}$ (red dot), spanned by the tangent vector $|\partial_j C(\vec{\theta})\rangle$ (blue arrow).  Figure and caption adapted from Ref.~\cite{Funcke:2021aps}.} 
    \label{fig_tangentspace}
\end{figure} 

In the following, we briefly describe the key concept of the dimensional expressivity analysis. A quantum circuit $C(\vec{\theta})$ contains unitary gates that depend on parameters $\theta$ and, thus, can be understood as a map from a parameter space into the quantum device state space. Thus, one can define two manifolds: (i) a manifold $M$ of states $| C(\vec{\theta})\rangle$ that the quantum circuit can reach and (ii) a manifold $S$ of all physical states of the quantum device. 
In order to obtain a minimal and maximally expressive circuit, the circuit has to only generate physical states ($M\subseteq S$), the number of parameters has to be equal to ${\rm dim}(S)$, and the co-dimension of $M$, which is defined as ${\rm codim}(M) = {\rm dim}(S)-{\rm dim}(M)$, has to vanish. Using a hybrid quantum-classical algorithm (see Ref.~\cite{Funcke_2021} for details), the co-dimension can be computed by determining the tangent vectors $|\partial_j C(\vec{\theta})\rangle$ 
for a given parameter $\theta_j$ and by testing their linear independence (see Fig.~\ref{fig_tangentspace} for an illustration).
 Using an iterative procedure, one can identify the redundant parameters $\theta_k$, for which $|\partial_{k} C(\vec{\theta})\rangle$ 
is a linear combination of $|\partial_{j} C(\vec{\theta})\rangle$ with $j \neq k$, and remove them by setting them to a constant value. This procedure results in a minimal quantum circuit, which is maximally expressive if the number of remaining parameters is equal to ${\rm dim}(S)$. As an example, Ref.~\cite{Funcke_2021} applied the dimensional expressivity analysis to the commonly used \texttt{EfficientSU2} 2-local quantum circuit of Qiskit~\cite{Qiskit}, shown exemplarily for 3 qubits in Fig.~\ref{fig_qiskitsu2}. In this quantum circuit, the coloured unitary gates turn out to be redundant and thus can be removed from the circuit.

\subsection{How to deal with gauge fields and fermions?}
\label{sec:fields}

The Hamiltonian lattice formulation of the field theories relevant for the Standard Model comprise two major ingredients: fermionic matter and bosonic (gauge) fields. 
For implementing the fermionic degrees of freedom, most simulations so far have used a Jordan-Wigner transformation to translate them to qubits. While the Jordan-Wigner mapping associates each fermionic degree of freedom to a single qubit, it transforms originally local terms to long-range interactions in form of Pauli strings for models in (2+1) dimensions and beyond~\cite{Martinez2016,Kokail2018,Klco2018,atas2021su2,Atas:2022dqm}. More local encodings avoiding the long-range Pauli strings are known~\cite{Bravyi2002,Verstraete2005,Ball2005}, and there are recent efforts to develop new locality-preserving mappings that are suitable for quantum computing~\cite{Whitfield2016,Steudtner2019,Derby2021}. However, these come in general at the expense of requiring more qubits than a Jordan-Wigner approach.  

In order to deal with the gauge fields, a variety of methods are available. Early works on digital quantum computing mainly focused on gauge models in (1+1) dimensions on lattices with open boundaries~\cite{Muschik2016,Kokail2018}, for which the gauge degrees of freedom can be completely integrated out~\cite{Sala2018}. This allows for obtaining a formulation directly on the gauge-invariant subspace of the theory. In higher dimensions, the gauge fields can no longer fully eliminated, and the infinite Hilbert spaces associated with the gauge fields have to be truncated to a finite dimension. Several approaches for such a truncation have been proposed in the literature. Using the basis in which the electric field energy is diagonal, the model can be truncated in the irreducible representations of the gauge group, while preserving the gauge symmetry~\cite{Zohar2015}. However, when using this approach, one is typically restricted to regimes where the (color) electric field is small, and one cannot explore the weak-coupling regime. Alternatively, one can work in the magnetic basis, which renders the magnetic term of the Hamiltonian diagonal, and truncate this basis at a finite number of elements~\cite{Haase2020,Bauer2021}. Similar to the truncation in the irreducible representations, this approach works best in regimes where one is close to an eigenstate of the magnetic part of the Hamiltonian and fails at strong coupling. Another approach that has been put forward is to digitize the gauge degrees of freedom by taking discrete subgroups or subsets of the gauge group~\cite{Gustafson2022,Hartung2022}. In addition, especially for non-Abelian theories, loop-string-hadron formulations have been suggested, which allow for solving the non-Abelian gauge constraints locally and only impose an Abelian constraint on the gauge links~\cite{Raychowdhury2020a,Raychowdhury2020,Kadam2022}. Moreover, quantum link models~\cite{Orland1990,Chandrasekharan1997} provide formulations of gauge theories, where the Hilbert space dimensions are by construction finite dimensional at the expense of an extra dimension which has to be taken to infinity to obtain the continuum limit~\cite{Brower1999,Brower2004}. For more details and references, see the review in Ref.~\cite{Bauer:2022hpo}.

\subsection{Which field theories have already been simulated on quantum computers?}

The ambitious goal of quantum computing the Standard Model is still far away, given the sizes and noise levels of current NISQ devices. Still, first quantum computations of (1+1)-dimensional gauge theories have already been accomplished, showing some of the relevant characteristics of the SM, including phase transitions~\cite{Martinez2016, Yang2020,Mil_2020,A_Rahman_2021,Zhou:2021kdl,Ciavarella:2021lel, Ciavarella:2021nmj, Klco:2019evd, Klco:2018kyo,deJong:2021wsd, Kokail2018, Schweizer:2019, mildenberger_probing_2022, atas2021su2, Atas:2022dqm, Farrell:2022wyt, Farrell:2022vyh,Huffman:2021gsi,Lumia:2021tpu}. In addition, resource-efficient formulations of gauge theories for quantum computations have been developed (see, e.g., Refs.~\cite{Banuls2017,Celi_2020,Kaplan2020,Haase2020,Paulson2020,armon2021photonmediated}). To perform lattice field theory computations on NISQ devices, one can use hybrid quantum-classical algorithms, such as VQE (see Sec.~\ref{sec:VQE}) and VQD (see Sec.~\ref{sec:VQD}).
These algorithms have already found numerous applications for studying benchmark models of particle physics in (1+1) dimensions~(see, e.g., Refs.~\cite{Banuls2020,klco2021standard, zohar2021quantum} for reviews). 

\subsubsection{Example: computing hadron masses of (1+1)-dimensional SU(2) gauge theory}
One example is the recently accomplished implementation of a non-Abelian gauge theory with both gauge and matter fields on a quantum computer~\cite{atas2021su2}. As shown in Fig.~\ref{fig:SU(2)-SU(3)-QC}(a), the hadron masses of the (1+1)-dimensional SU(2) gauge theory were computed using VQE (see Sec.~\ref{sec:VQE}) and the IBM-Q Casablanca quantum processor. The corresponding lattice Hamiltonian reads 
\begin{equation}
		\hat{H}_{\rm SU(2)_{(1+1)D}} = \frac{1}{2a} \sum_{n=1}^{N-1} \left( \hat{\phi}_{n}^\dagger \hat{U}_{n} \hat{\phi}_{n+1} + \operatorname{h.c.}\right) +m \sum_{n=1}^{N} (-1)^{n} \hat{\phi}_{n}^\dagger \hat{\phi}_{n}  + \frac{a g^{2}}{2} \sum_{n=1}^{N-1} \hat{\boldsymbol{L}}_{n}^{2},
		\label{eq:KSham}
\end{equation}
where $N$ is the number of lattice sites with spacing $a$, $g$ is the gauge coupling, $m$ is the mass of the ``quark'' fermion, $\hat{\phi}_n = \left( \hat{\phi}^1_n, \hat{\phi}^2_n \right)^T$ is the staggered ``quark'' fermion field at site $n$ with two color components (red and green), $\hat{U}_n$ is the gauge link connecting sites $n$ and $n+1$, and $\hat{\boldsymbol{L}}_{n}^{2}=\sum_{a}\hat{L}_{n}^{a}\hat{L}_{n}^{a}=\sum_{a}\hat{R}_{n}^{a}\hat{R}_{n}^{a}$ is the color electric field energy, where $\hat{L}_{n}^{a}$  and $\hat{R}_{n}^{a}$ (with $a=x,y,z$) are the left and right color electric field components on the link $n$.
For the non-Abelian SU(2) gauge group, the right and left color electric field are different and are related via the adjoint representation $\hat{R}_{n}^{a}=\sum_b(\hat{U}_n^{\text{adj}})_{ab}\hat{L}_{n}^{b}$, where $(\hat{U}_{n}^{\text{adj}})_{ab}=2\mathrm{Tr}\left[ \hat{U}_{n}\hat{T}^{a}\hat{U}_{n}^{\dagger}\hat{T}^{b}\right]$,  $\hat{T}^{a}=\hat{\sigma}^{a}/2$ are the three generators of the SU(2) algebra, and $\hat{\sigma}^{a}$ are the Pauli matrices. In Fig.~\ref{fig:SU(2)-SU(3)-QC}(a), the mass $M_{\text{b}} = E_\text{b} - E_\text{v}$ of the lightest baryon is shown, which is defined as the energy gap between the lowest baryon state $ E_\text{b}$ and the vacuum state $E_\text{v}$. In order to obtain these results, Ref.~\cite{atas2021su2} prepared both states on the quantum hardware, for the exemplary parameters $N = 4$, $\tilde{m} \equiv am= 1$, and $x\equiv 1/(ag)^2 \in [0,5]$.

\subsubsection{Example: performing real-time evolution for (1+1)-dimensional SU(3) gauge theory}

\begin{figure}[!t]
    \centering
    \subfloat[\centering VQE results for (1+1)-dim.\ SU(2) + matter~\cite{atas2021su2}.]{{\includegraphics[width=.43\linewidth]{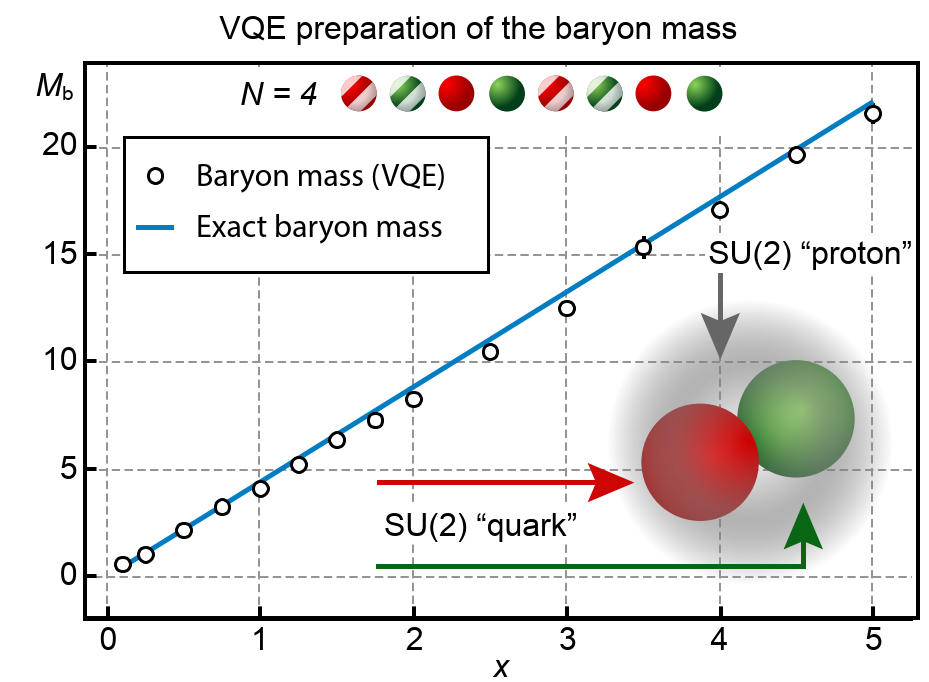} }}
    \qquad
    \subfloat[\centering Time evolution for (1+1)-dim.\ pure SU(3)~\cite{Ciavarella:2021nmj}.]{{\includegraphics[width=.45\linewidth]{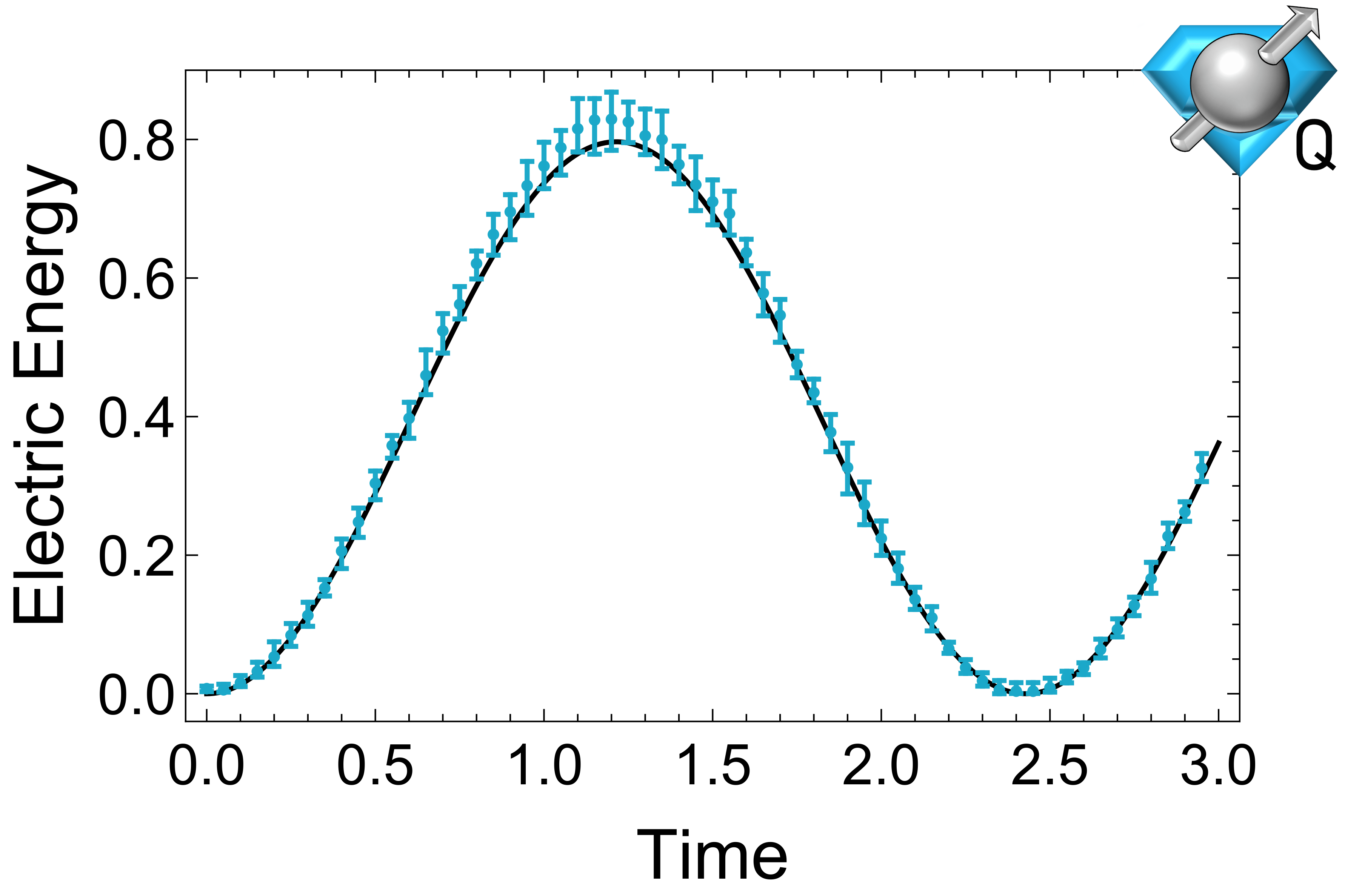} }}
    \caption{(a) VQE calculation of the baryon mass, for $N=4$ lattice sites, a ``quark'' mass of $\tilde{m}=am=1$, and a range of values for the inverse coupling constant $x=1/(ag)^2\in [0,5]$. The baryon is an SU(2)-“proton” (see inset), and the error bars are smaller than the markers. Figure reprinted with permission from Ref.~\cite{atas2021su2}, \url{https://doi.org/10.1038/s41467-021-26825-4}, under the \href{https://creativecommons.org/licenses/by/4.0/}{Creative Commons Attribution 4.0 International License}. (b) VQE results for the energy in the electric field of the one-plaquette SU(3) system evolved according to the Hamiltonian in Eq.~\eqref{eq:SU(3)}. The data points correspond to the average value and the maximal extent of 68\% binomial confidence intervals across four implementations on IBM-Q’s Athens quantum processor.  Figure reprinted with permission from Ref.~\cite{Ciavarella:2021nmj}, \url{https://doi.org/10.1103/PhysRevD.103.094501}, copyright (2021) by the Americal Physical Society.}
    \label{fig:SU(2)-SU(3)-QC}
\end{figure}

Going from SU(2) to SU(3) gauge theory, Ref.~\cite{Ciavarella:2021nmj} recently accomplished the first quantum computation of pure (1+1)-dimensional SU(3) gauge theory dynamics. Instead of discussing the original lattice Hamiltonian considered in Ref.~\cite{Ciavarella:2021nmj}, we would like to now give an example of the final Hamiltonian used for the actual quantum computation, after applying the methods discussed in Sec.~\ref{sec:fields}. Using the lowest non-trivial truncation in the color parity basis, the original lattice Hamiltonian of the SU(3) gauge theory was simplified to~\cite{Ciavarella:2021nmj}
\begin{equation}
	\hat{H}_{\rm trunc.SU(3)_{(1+1)D}} = \left(\frac{4}{3}g^2 + \frac{11}{4g^2} \right) \hat{\mathds{1}} + \left(-\frac{4}{3}g^2 + \frac{1}{4g^2} \right)\hat{Z} - \frac{1}{\sqrt{2} g^2}  \hat{X},
	\label{eq:SU(3)}
\end{equation}
where $g$ is the gauge coupling and $\hat{X}$ and $\hat{Z}$ are the first and third Pauli operators, respectively. 
Figure~\ref{fig:SU(2)-SU(3)-QC}(b) shows the results of performing the associated real-time evolution on IBM-Q's Athens quantum processor with $g=1$, starting in the electric vacuum. While these results have been obtained for pure SU(3) gauge theory without fermions, we note that there has very recently been the first quantum computation of real-time evolution of tetraquark physics in SU(3) gauge theory plus fermionic matter, i.e., (1+1)-dimensional ``QCD''~\cite{Atas:2022dqm}. Moreover, vacuum dynamics and $\beta$-decay for one- and two-flavor QCD in (1+1)-dimensions have been simulated on NISQ devices~\cite{Farrell:2022wyt,Farrell:2022vyh}.

\subsection{Outlook: combining quantum computations with classical MCMC computations}

\begin{figure}[!t]
    \centering
    \includegraphics[width=0.65\linewidth]{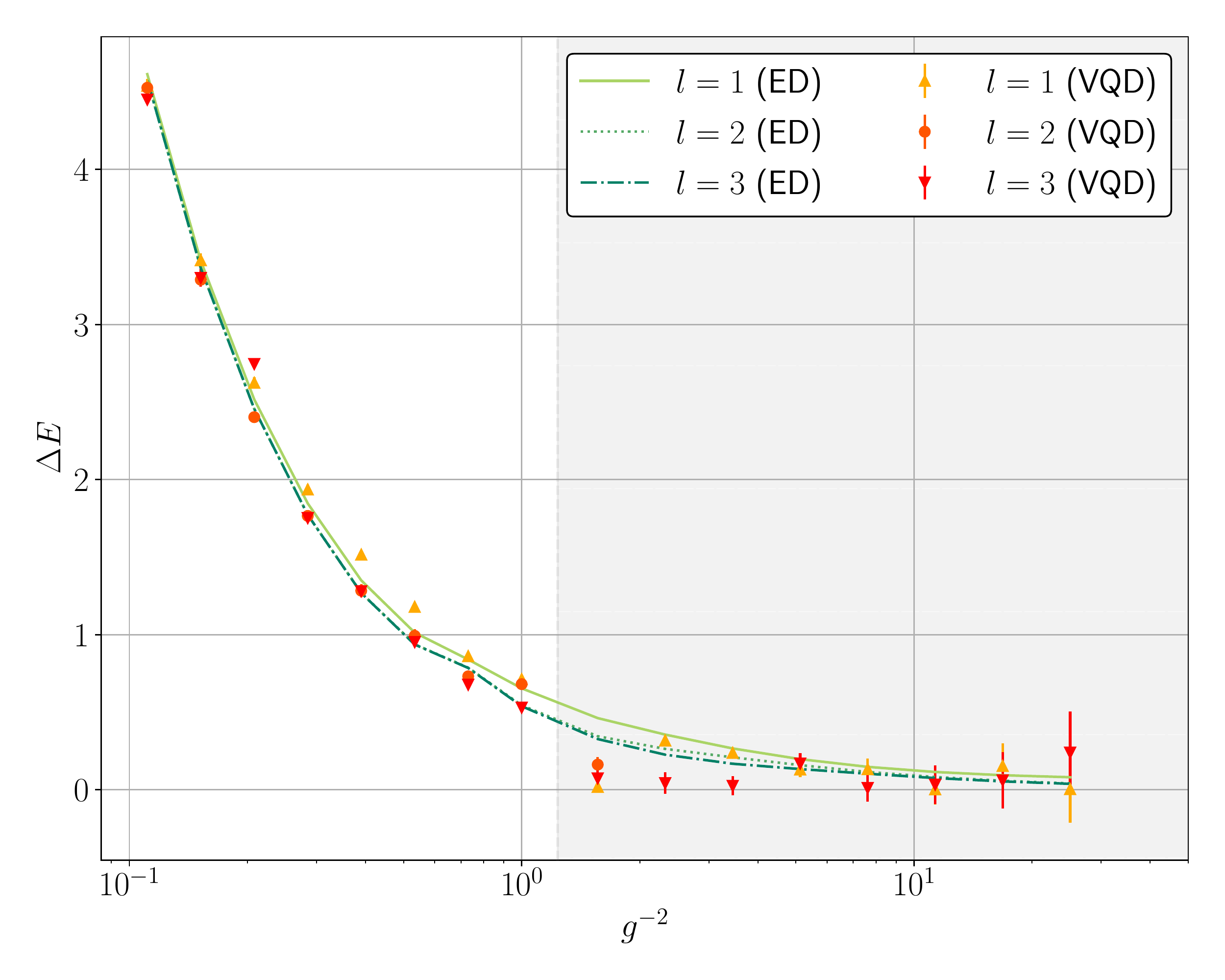}
    \caption{Classically simulated VQD results of spectral gap $\Delta E = E_1-E_0$ for (2+1)-dimensional QED in the electric basis, as a function of the gauge coupling $g$ at vanishing fermion mass, $m=0$. Both the VQD results (dots) and the exact diagonalization (ED) results (lines) are shown for truncation levels up to $l=3$. The shaded area corresponds to the region where the results are not precise enough to estimate the gap reliably. Figure and caption adapted from Ref.~\cite{Clemente_Strategies_2022}.}
    \label{fig:VQD}
\end{figure}

In this section, we would like to discuss an example of \textit{combining} quantum computations with classical MCMC computations in the future. As proposed in Ref.~\cite{Clemente_Strategies_2022}, one may combine small-scale quantum computations with large-scale MCMC computations, in order to overcome the MCMC problem of critical slowing down (see Sec.~\ref{sec:classical}). In contrast to MCMC methods, quantum computing does not face any obstacles when investigating small lattice spacings. Thus, one may match the results of short-distance quantities obtained from quantum computations with the ones coming from MCMC simulations in the strong and intermediate coupling regime. This matching can then be used, e.g., to compute the $\Lambda$-parameter~\cite{Clemente_Strategies_2022}. In order to set the physical value of the lattice spacing, one needs to compute an observable, such as the spectral gap $\Delta=E_1-E_0$. Figure~\ref{fig:VQD} shows the results for the spectral gap of (2+1)-dimensional QED for truncation levels up to $l=3$, using the VQD algorithm (see Sec.~\ref{sec:VQD}) and Qiskit's \texttt{EfficientSU2} quantum circuit (shown exemplarily for 3 qubits and 2 layers in Fig.~\ref{fig_qiskitsu2}) with up to 5 layers. The results in Fig.~\ref{fig:VQD} have been obtained using a classical simulator of the quantum hardware without noise and with an infinite number of shots, in order to test the feasibility of the method for a small number of qubits, and the method can be implemented on quantum hardware in the future. 

Other avenues of combining small-scale quantum computations with large-scale MCMC computations have been proposed, e.g., for addressing the MCMC problem of interpolator optimization~\cite{Avkhadiev:2019niu, Avkhadiev:2022ttx}. In general, for computations that combine lattice results from the Hamiltonian and Lagrangian formulations, several challenges need to be addressed. First, lattice field theories are usually expressed in the Lagrangian formulation, and one needs to derive and optimize the corresponding Hamiltonian formulation (see, e.g., Refs.~\cite{kan2021investigating, Carena:2022kpg, Carena:2021ltu,Cohen:2021imf}).
Second, one needs to match the different bare parameters and observables between these two formulations (see, e.g., Refs.~\cite{Loan:2002ej,Byrnes:2003gg, Carena:2022hpz, Funcke:2022opx}). Third, one needs to implement the same lattice fermion formulation, but lattice computations in the Hamiltonian formulation have so far mainly focused on staggered fermions. Recently, there has been progress with the implementation of Wilson fermions in the Hamiltonian formulation~(see, e.g., Refs.~\cite{Zache_2018,Mazzola_2021}) and with determining the resulting mass shift~\cite{Angelides:2022pah}.

\section{Summary and outlook: where do we stand, where will we go?}
\label{sec:summary}

\subsection{Short summary: quantum hardware and algorithms}

Quantum computing offers the prospect to overcome challenges of MCMC methods, including the sign problem and the problem of critical slowing down. On the hardware side, digital gate-based quantum computers with $\mathcal{O}(10-100)$ noisy qubits are currently available. On the algorithms side, first resource-efficient quantum algorithms for gauge theories in (1+1) and (2+1) dimensions have been developed, and first proof-of-concept quantum computations of gauge theories in (1+1) dimensions have been accomplished. Thus, the research field of quantum computing for lattice field theory is still in its very early stages of development. The path towards quantum computations of lattice field theories in (3+1) dimensions, in particular Lattice QCD, requires numerous steps of quantum algorithm and hardware improvement. 

\subsection{Discussion and outlook: the future of quantum computing}

\begin{figure}[!htb]
    \centering
\includegraphics[width=.5\linewidth]{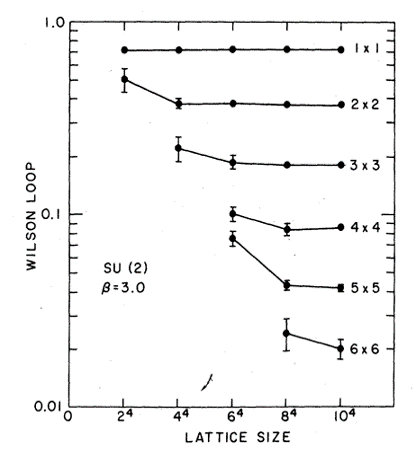}
\caption{First lattice field theory computation using classical MC methods. Wilson loops of pure SU(2) gauge fields at $\beta=3$ as a function of lattice size. Figure reprinted with permission from Ref.~\cite{Creutz:1980zw}, \url{https://doi.org/10.1103/PhysRevD.21.2308}, copyright (1980) by the Americal Physical Society.}
    \label{fig:Wilson_loops_1980}
\end{figure}

The field of quantum computing recently entered the era of noisy, intermediate-scale quantum devices, allowing for first computations of lattice gauge theories on lattice sizes comparable to the ones in the pioneering work by Creutz~\cite{Creutz:1980zw} more than 40 years ago (see Fig.~\ref{fig:Wilson_loops_1980}). The investigation of lattice field theories with classical MCMC algorithms has seen tremendous advances in these 40 years, currently allowing the simulation of QCD with physical values of the quark masses for the first two generations of quarks. To reach the current era of large-scale, high-precision computations on today's supercomputers, which allows the computation of the light QCD spectrum and much more, both the classical algorithms and the classical hardware had to be improved to a level that might have been considered impossible from the perspective of the 1980's.

Thus, one may ask: can the field of quantum computing for lattice field theory develop similarly to the field of classical computing for lattice field theory, reaching a similar level of precision in the upcoming decades? Of course, this question cannot be answered, because such an answer strongly depends on the future development of quantum hardware and algorithms, which are both unknown. However, one can make a few rough estimates regarding the requirements of quantum computing for Lattice QCD. The spatial lattice volumes that can currently be simulated classically are large, up to $96^3$, and would require $\mathcal{O}(10^7-10^8)$ qubits on the quantum computing side~\cite{Banuls2019a}. Considering  fault-tolerant logical qubits, this number would get multiplied by a factor of $\sim 1000$, as explained in Sec.~\ref{sec:correction}. The quantum computing roadmap of various companies suggests that fault-tolerant quantum computing might become reality in the current or next decade (see, e.g.. Ref.~\cite{IBMQ-Roadmpap2020,Google-Roadmpap2021}). Following these roadmaps and extrapolating them in time (which poses various challenges for quantum hardware and error correction), a quantum computation of Lattice QCD with spatial lattice volumes of $96^3$ might become feasible in the 2040s or 2050s. This time period would be long but comparable to the time since the first days of classical lattice field theory computations. 

At this point, we wish to emphasize that for quantum computing to become useful for lattice field theory computations, spatial lattice volumes as large as $96^3$ are not required. For simulating sign-problem-afflicted regimes that can neither be studied with MCMC nor with classical methods beyond MCMC, such as TN methods, it would be sufficient to perform quantum computations with much smaller lattice volumes. Examples of outperforming the best current classical algorithms in sign-problem-afflicted regimes have already been provided, e.g., in the context of (1+1)-dimensional lattice field theories, as explained in Sec.~\ref{sec:analog}.

\subsection{Discussion and outlook: the race between quantum and classical computing}

In the previous section, we brought up a comparison between the history of classical computing and the possible future of quantum computing for lattice field theory. This comparison raises an important conceptual question: if the classical algorithms and hardware have progressed so tremendously in the past, might they develop quickly enough in the future to catch up with any progress in quantum computing? There have been several examples in the history of quantum computing, where a quantum advantage turned into a ``classical advantage'' after the classical algorithms had been improved. For example, Ref.~\cite{Arute2019} demonstrated a quantum computation on 53 qubits taking 200s, substantially shorter than the run time of 10,000y of the classical algorithm. Due to a loophole in the classical algorithm, the run time was later reduced to a few days, also when slightly increasing the qubit number~\cite{IBMQ-Supremacy2019}, and was further reduced to 304s in Ref.~\cite{Liu:2021xar}, thus comparable to the run time of the quantum computation.

Would it be possible that classical algorithms for lattice field theory, including MCMC-based methods, TN methods, machine-learning methods, and other methods to overcome or mitigate the sign problem, might advance quickly enough in the upcoming decades that quantum computing would not be needed anymore? This is again a question that cannot be answered, because the answer strongly depends on the future development of classical hardware and algorithms, which are both unknown. Even when classical and quantum computations can compete, quantum computing could still give advantages, e.g., in specific parameter regimes of the lattice field theory, or more generally, e.g., due to lower power consumption. Crucially, whenever one encounters an exponentially hard classical problem, a small quantum step corresponds to a giant classical leap. For example, to simulate out-of-equilibrium dynamics, the errors of the best known classical algorithms grow exponentially in time, as discussed in Sec.~\ref{sec:analog}. For highly entangled quantum systems, we expect that quantum computing will be able to outperform classical computing for lattice field theories in (3+1) dimensions in the future, once sufficient resources will be available.

\acknowledgments
L.F.\ is partially supported by the U.S.\ Department of Energy, Office of Science, National Quantum Information Science Research Centers, Co-design Center for Quantum Advantage (C$^2$QA) under contract number DE-SC0012704, by the DOE QuantiSED Consortium under subcontract number 675352, by the National Science Foundation under Cooperative Agreement PHY-2019786 (The NSF AI Institute for Artificial Intelligence and Fundamental Interactions, \url{http://iaifi.org/}), and by the U.S.\ Department of Energy, Office of Science, Office of Nuclear Physics under grant contract numbers DE-SC0011090 and DE-SC0021006.
S.K.\ acknowledges financial support from the Cyprus Research and Innovation Foundation under projects ``Future-proofing Scientific Applications for the Supercomputers of Tomorrow (FAST)'', contract number COMPLEMENTARY/0916/0048, and “Quantum Computing for Lattice Gauge Theories (QC4LGT)”, contract number EXCELLENCE/0421/0019.

\bibliographystyle{h-physrev}
\bibliography{ref}

\end{document}